\documentclass[letterpaper, 10 pt, conference]{ieeeconf}  % Comment this line out if you need a4paper

\IEEEoverridecommandlockouts                              % This command is only needed if 
\usepackage[nolist,nohyperlinks]{acronym}
\usepackage{amsmath}
\usepackage{graphicx}
\usepackage{hyperref}
\usepackage{booktabs}
\usepackage{lipsum}
\usepackage{footnote}
\usepackage{mathrsfs}
\makesavenoteenv{tabular}
\usepackage{makecell}
\usepackage{colortbl} % Include the colortbl package
\definecolor{lightgray}{gray}{0.9} % Define a custom color


\title{\LARGE \bf
Leveraging Cognitive States for Adaptive Scaffolding of Understanding in Explanatory Tasks in HRI
}
\author{André Groß$^{1,2*}$, Birte Richter$^{1,2}$, Bjarne Thomzik$^{1,2}$ and Britta Wrede$^{1,2}$% <-this % stops a space
\thanks{$^{1}$Medical Assistance Systems, Medical School OWL and $^{2}$Center for Cognitive Interaction Technology, CITEC from Bielefeld University, Germany. $^{*}$Corresponding Author: André Groß \tt\small agross@techfak.uni-bielefeld.de.}%
\thanks{This research was funded by the Deutsche Forschungsgemeinschaft (DFG, German Research Foundation): TRR 318/1 2021-438445824 “Constructing Explainability”.}%
}

\begin{document}

%%%%%%%%%% acronym list
\begin{acronym}[list of abbreviations]
\acro{hri}[HRI]{\textit{Human-Robot Interaction}}
\acro{xai}[XAI]{\textit{Explainable Artificial Intelligence}}
\acro{ros}[ROS]{\textit{Robot-Operating-System}}
\acro{scxml}[SCXML]{\textit{State Chart eXtensible-Markup-Language}}
\acro{its}[ITS]{\textit{Intelligent Tutoring Systems}}
\acro{rl}[RL]{\textit{Reinforcement-Learning}}
\acro{aoi}[AOI]{\textit{Area of Interest}}
\acro{vfoa}[VFoA]{\textit{Visual Focus of Attention}}
\acro{lme}[LME]{\textit{Linear Mixed-Effects Model}}
\acro{seq}[SEQ]{\textit{Single Ease Question}}
\acro{stm}[STM]{\textit{Short-Term Memory}}
\end{acronym}
%%%%%%%%%% acronym list end

\maketitle
\thispagestyle{empty}
\pagestyle{empty}

%%%%%%%%%%%%%%%%%%%%%%%%%%%%%%%%%%%%%%%%%%%%%%%%%%%%%%%%%%%%%%%%%%%%%%%%%%%%%%%%
\begin{abstract}
Understanding how scaffolding strategies influence human understanding in human-robot interaction is important for developing effective assistive systems.
This empirical study investigates linguistic scaffolding strategies based on negation as an important means that de-biases the user from potential errors but increases processing costs and hesitations as a means to ameliorate processing costs.
In an adaptive strategy, the user state with respect to the current state of understanding and processing capacity was estimated via a scoring scheme based on task performance, prior scaffolding strategy, and current eye gaze behavior.
In the study, the adaptive strategy of providing negations and hesitations was compared with a non-adaptive strategy of providing only affirmations.
The adaptive scaffolding strategy was generated using the computational model SHIFT.
Our findings indicate that using adaptive scaffolding strategies with SHIFT tends to (1) increased processing costs, as reflected in longer reaction times, but (2) improved task understanding, evidenced by a lower error rate of almost 23\%.
We assessed the efficiency of SHIFT's selected scaffolding strategies across different cognitive states, finding that in three out of five states, the error rate was lower compared to the baseline condition.
We discuss how these results align with the assumptions of the SHIFT model and highlight areas for refinement.
Moreover, we demonstrate how scaffolding strategies, such as negation and hesitation, contribute to more effective human-robot explanatory dialogues.
\end{abstract}
%%%%%%%%%%%%%%%%%%%%%%%%%%%%%%%%%%%%%%%%%%%%%%%%%%%%%%%%%%%%%%%%%%%%%%%%%%%%%%%%

\section{Introduction}

In the growing field of social robotics, robots are increasingly being designed to assist people in their everyday lives.
From educational support to collaborative tasks, these systems aim to enhance human capabilities by interacting and guiding \cite{ramachandran2019personalized, wang2018facilitating, clement2024arxiv}.
Social robots are expected to engage in meaningful, goal-driven conversations and adjust to the users' needs rather than only executing commands \cite{almasri2019intelligent}.
This ability to dynamically support human learning and problem-solving is crucial in settings where robots take on the role of tutors or assistants.
In human-robot communication research, robots have already been used successfully as explainers~\cite{matarese2023ex}.
However, their responses are often static, lacking the ability to adapt conversations to the specific needs of the human \cite{fong2003collaboration}.
This highlights the challenge in \ac{hri} to establish a natural dialogue and enable the robot to respond to the user's personal needs.
In education and skill acquisition, human tutors employ scaffolding -- a process of gradually adjusting support levels based on the learner’s progress \cite{vollmer2010developing}.
Effective scaffolding ensures that learners receive the right level of assistance at the right time, fostering independence over time.
For robots to function effectively as instructors or collaborative partners, they must also incorporate this adaptive scaffolding approach, dynamically adjusting their explanations and guidance in response to learning behaviors \cite{leyzberg2014personalizing}.
The explainer needs to be able to interpret the learner's cues during communication, while continuously monitoring the learner's progress \cite{rohlfing2020explanation}.
Recognizing these signals and checking progress is the challenge in social robotics and tutoring settings in general.
To address this challenge, we present a study that explores the use of verbal scaffolding strategies to enhance human attention and understanding during task-solving.
In a user study, we demonstrate: the \textbf{measurement of human attention and task understanding} as the human cognitive state, the \textbf{impact of different scaffolding strategies} on human processing capacity and task performance, and the \textbf{adaptive selection of scaffolding strategies} by a computational model based on human monitoring, improving the learning process.

\begin{figure}[!t]
    \centering
    \includegraphics[width=\linewidth]{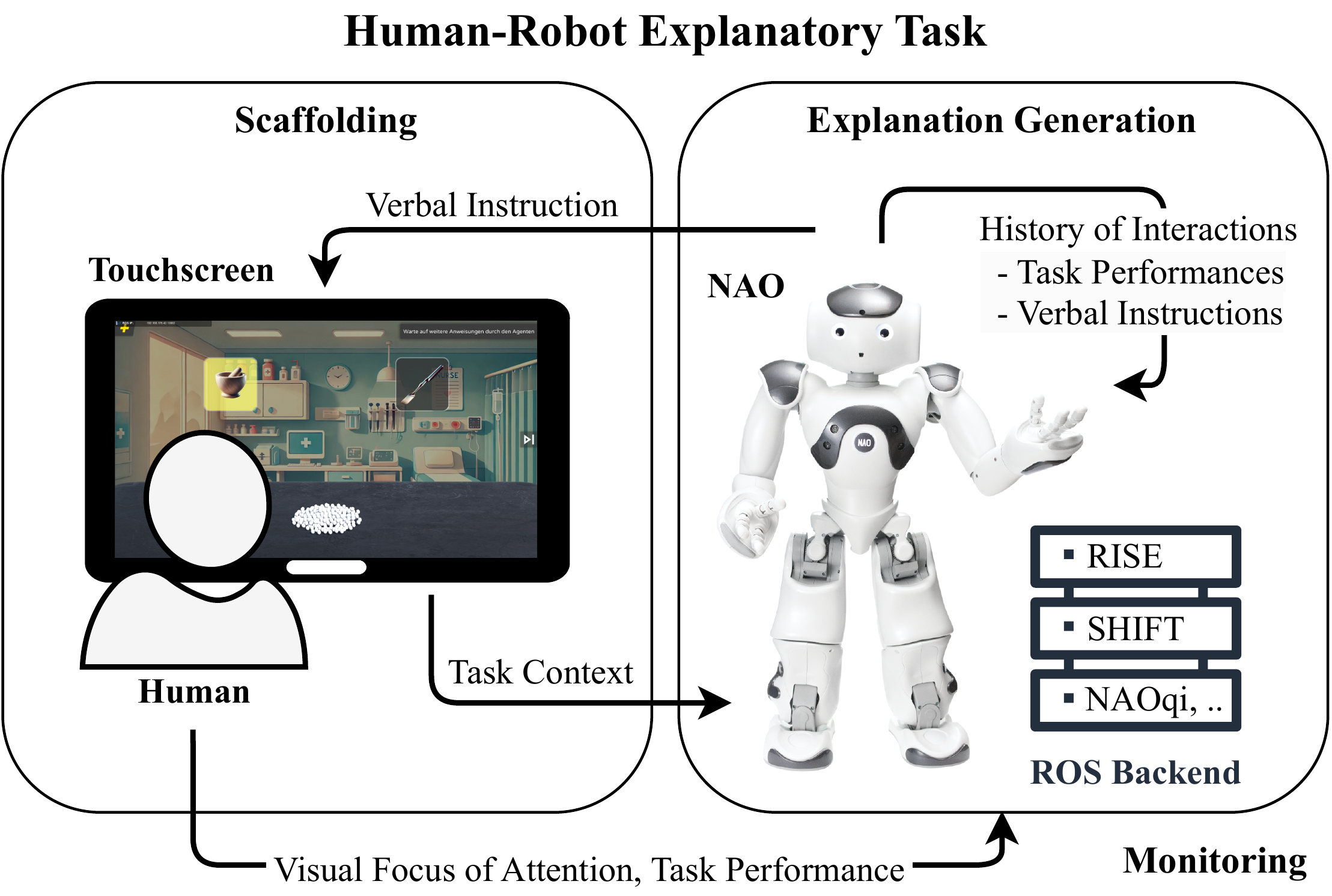}
    \caption{\acl{hri} study design: The NAO robot provides verbal instructions to guide humans in completing tasks on a touchscreen. The explanation generation is based on human monitoring.}
    \label{fig:study_setting}
    \vspace{-20pt}
\end{figure}

\section{Related Work}

\subsection{Intelligent Tutoring Systems}

In educational settings, the challenge of providing adaptive guidance is tackled by \ac{its}, which are structured around three core models: the pedagogical model, the student model, and the domain model \cite{almasri2019intelligent}.
The pedagogical model incorporates educational principles and strategies, determining how a topic should be taught based on established teaching methodologies.
The student model represents the system's knowledge of the learner’s current understanding and progress.
In robotics terms, this corresponds to the partner model, which enables the system to adapt its support to the individual needs of the user.
The domain model defines the structure and rules of the context (explanandum) being taught.
It provides the necessary constraints and guidelines for understanding the context or solving tasks, ensuring that explanations follow a logical sequence relevant to the topic.
The primary goal of \ac{its} is to dynamically adjust the level of guidance based on the pedagogical model and the learner's progress as represented in the student model, while respecting the rules defined in the domain model.
Different \ac{its} implementations use varying modalities of support.
Some systems adapt scaffolding strategies, ranging from minimal intervention (such as brief pauses for self-recovery) to fully interactive tutorial sessions, for example in math \cite{ramachandran2019personalized}.
Other systems use a rule-based domain model that optimizes the order of explanations.
In information technology education, a \ac{its} may structure explanations progressively, introducing concepts such as “file” before “database”, ensuring a logical increase in complexity. \cite{wang2018reinforcement}.
While these systems focus on adjusting the level and timing of explanations, they often overlook the use of different verbal scaffolding strategies.
Most approaches modify the amount of information provided, but do not consider how it is provided.
Therefore, we investigate how different verbal scaffolding strategies influence human task understanding.

\subsection{Scaffolding in Human-Robot Interaction}

We already know from human-human communication that specific linguistic strategies, verbal utterances, can be used to guide users effectively.
Two such strategies are negation and hesitation, which have different effects on human cognitive processing.
Negating serves as corrective feedback, signalling mistakes while maintaining engagement.
Hesitations introduce pauses in explanation, allowing learners time to process information, encourage self-correction and promote reflection without overwhelming the user.
The use of such verbal expressions is also part of \ac{hri}.
A separate area of research is to achieve natural dialogue with optimal verbal explanations.
In a previous user study \cite{gros_scaffolding_nodate}, we have shown that the use of contrastive explanations, in terms of negation, show different effects on human's processing. 
The study reveals that a negation is a cognitively more demanding strategy for human processing, measured in terms of reaction time, than an affirmation. 
This higher level of cognitive effort translated into a better task performance, as measured by movement accuracy. 
Regarding the use of hesitations, we have shown in previous studies that hesitations as a scaffolding strategy in human-robot tutoring successfully regained the human's attention during distraction and yielded better retention \cite{betz2018interactive, richter2021attention}.
Indeed, in an EEG analysis, we could verify that hesitations change neural brain activity significantly in a \ac{hri} \cite{richter2023eeg}.
These studies highlight the potential and benefits of different verbal scaffolding strategies, but do not address an adaptive approach for robots to determine which strategy to use.

\subsection{Cognitive Modeling, Partner-Model} 

To adaptively select effective scaffolding strategies based on the human's current state, the robot must have an understanding of the human's internal processes.
This requires a formal representation of the human's contextual understanding, known as the partner model \cite{buschmeier2014dynamic}.
Adaptive responses to changes in the partner model rely on interpreting snapshots of the model as representations of the human's cognitive states~\cite{buschmeier2014dynamic}.
These snapshots must be continuously monitored and used to adjust the system in real time.
Various social cues, such as attention, task performance and interaction history, can be used to infer cognitive states within a model.
In~\cite{gross2025shift}, we introduced \textit{SHIFT}, a domain-independent approach for adaptive scaffolding in robotic explanation generation to support task guidance in \ac{hri}. 
Our approach integrates interdisciplinary research findings into a computational model based on a pre-configured scoring system. 
\textit{SHIFT} represents the human cognitive state using six observable states within the human partner model. 
A \ac{rl} approach enables adaptation to individual deviations from the norm.
However, limited research has explored how the selection of different scaffolding strategies, adapted to the user's cognitive state, affects task understanding.

\subsection{Research Questions and Hypotheses} \label{sec:research_questions}

This work contributes to the field by integrating cognitive and social factors into a study of human-robot interaction that goes beyond static experimental designs and aims for greater real-world applicability.
Building on the study setting from previous work \cite{gros_scaffolding_nodate}, we demonstrate its expansion and incorporate a model for the adaptive generation of explanatory strategies \cite{gross2025shift}.
This work focuses on the following research questions and hypotheses:

\begin{itemize} 
    \item[1)] How is the adaptive generation of scaffolding strategies based on the human cognitive state influencing human task performance in robot-assisted interactions compared to affirmations?
    \begin{itemize}
        \item[$H_1$)] The use of different scaffolding strategies provided by \textit{SHIFT} increases human processing costs, as indicated by longer reaction times, compared to affirmations.
        \item[$H_2$)] The use of \textit{SHIFT} fosters task understanding, as evidenced by a reduction in task-solving errors, compared to affirmations.
    \end{itemize}
    \item[2)] How effectively does \textit{SHIFT} select appropriate scaffolding strategies based on observed human behavior, measured by the number of failures in each cognitive state, including metrics of task awareness, processing capacity, and interaction history?
\end{itemize}

\section{Method}

\begin{table*}[!t]
\renewcommand{\arraystretch}{1.3}
\caption{Overview of tasks with two possible actions for completion, including the reasons for each action.}
\centering
\begin{tabular}{rllllll}
  \toprule
  & Task        & Action I            & Action II         & Reason I         & Reason II              & Visual Feedback \\ 
  \hline 
  1 & Pill      & crush               & break             & Risk of choking       & Sensitive stomach           & Changes to mesh \\
  2 & Bottle    & shake               & swirl             & Respiratory disease   & Gastrointestinal problems   & Liquid color \\
  3 & Injection & draw slowly         & draw quickly      & Sensitive tissue      & Allergic reaction           & Liquid color \\
  4 & Pavement  & spread horizontally & spread vertically & Longitudinal injury   & Transverse injury           & Line color \\
  5 & Salve     & rotate              & press             & Small wound           & Large wound                 & Numerator display \\
   \bottomrule
\end{tabular}
\label{tab:tasks}
\end{table*}

\begin{figure*}[!t]
    \centering
    \includegraphics[width=\linewidth]{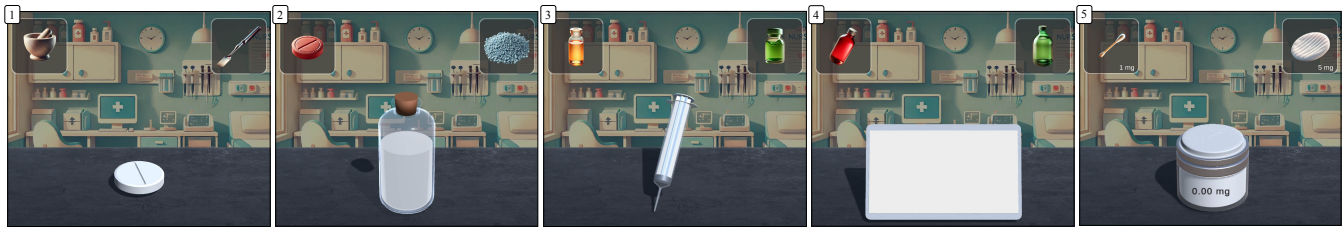}
    \caption{Visualization of tasks with three target stimuli: the main object to be manipulated in the center and two objects (tool 1, tool 2) visible in the upper corners and associated with corresponding actions (action 1, action 2) from \autoref{tab:tasks}.}
    \label{fig:tasks}
    \vspace{-18pt}
\end{figure*}

This paper presents a \ac{hri} study conducted in German, where the robot NAO \cite{naogouaillier2009mechatronic} assists humans in solving tasks on a touchscreen using different verbal scaffolding strategies (\autoref{fig:study_setting}).
The study investigates the effects of these strategies on task understanding.

\subsection{Experimental Conditions}

A between-subjects design was used to compare two experimental conditions.
In the baseline condition (\textit{BL}), participants received only affirmation-type explanations.
These explanations were consistently applied regardless of external observations.
In contrast, the adaptive condition (\textit{SHIFT}) provided participants with verbal scaffolding strategies that were selected in real time based on their cognitive state as determined by the monitoring of their social cues by our computational model \textit{SHIFT}.

\subsection{Participants}

A study with 34 participants was conducted.
Due to technical problems with either the eye tracking or the robot, four data sets were deemed invalid.
This left a total of 30 participants.
Participants were recruited from Bielefeld University and the University of Paderborn.
Non-students were also recruited via social media.
The participants were assigned to the experimental groups alternately.
The baseline group (\textit{BL}) consisted of 15 participants (8 female, 7 male) with an age range of $19-61$ years (mean $\overline{AGE}_\textit{BL} = 28$, $SD_{\textit{BL}} = 11.53$), while the adaptive group (\textit{SHIFT}) similarly included 15 participants (7 female, 8 male) with an age range of $20-66$ years (mean $\overline{AGE}_\textit{SHIFT} = 29$,  $SD_{\textit{SHIFT}} = 12.13$).
Both groups showed no significant difference in their mean score for technology affinity ($\overline{ATI}_\textit{BL} = 3.64$, $SD_{\textit{BL}} = 1.24$, $\overline{ATI}_\textit{SHIFT} = 4.02$, $SD_{\textit{SHIFT}} = 1.00$) and in their score for \ac{stm} ($\overline{STM}_\textit{BL, SHIFT} = 72.67$, $SD_{\textit{BL}} = 15.80$, $SD_{\textit{SHIFT}} = 11.00$).

\subsection{Tasks and Verbal Instructions}

This study investigates the impact of verbal scaffolding strategies on task understanding.
We designed five tasks for a touchscreen (\autoref{fig:tasks}), each involving three objects: a target object that has to be manipulated and two tools.
Each tool corresponds to a specific gesture that must be applied to the target object on the touchscreen.
Each task is divided into two subtasks: selection and interaction (\autoref{fig:task_flow}).
Participants first select the appropriate tool and then perform the appropriate gesture on the target object to complete the task.
Throughout the process, the robot provides verbal guidance (\autoref{tab:verbal_instructions}) to assist in tool selection.
To achieve this, the content of a verbal utterance is selected from a set of preconfigured sentences.

\subsection{Experimental Procedure}

\begin{table*}[!ht]
\renewcommand{\arraystretch}{1.3}
\caption{Types of verbal instruction structures with the original stimuli in German and translated examples in English.}
\centering
\begin{tabular}{rllll}
  \toprule
    & Human Cognitive State     & Instruction Structure                 & Stimuli, Example (GER)                                                        & Example (ENG)                         \\ 
  \hline 
  a & Engaged Observer          & Affirmation                           & Zerdrücke die Tablette                                                        & Crush the pill                        \\
  b & Engaged Misinterpreter    & Negation+Affirmation                  & \makecell[l]{Zerteile die Tablette \textbf{nicht}, \\ sondern zerdrücke sie}  & Do \textbf{not} break the pill, crush it        \\
  c & Distracted Misinterpreter & Negation                              & Zerteile die Tablette \textbf{nicht}                                          & Do \textbf{not} break the pill                  \\
  d & Overwhelmed Struggler     & Affirmation \& Hesitation             & \textbf{Mhm..} zerdrücke die Tablette                                         & \textbf{Mhm..} crush the pill                  \\
  e & Unfocused                 & Negation+Affirmation \& Hesitation    & \makecell[l]{Zerteile die Tablette \textbf{mhm..} \textbf{nicht}, \\ sondern zerdrücke sie}    & \textbf{Mhm..} do \textbf{not} break the pill, crush it  \\
  f & Uncertain                 & Negation \& Hesitation                & Zerteile die Tablette \textbf{mhm..} \textbf{nicht}                           & \textbf{Mhm..} do \textbf{not} break the pill            \\
   \bottomrule
\end{tabular}
\label{tab:verbal_instructions}
\end{table*}

\begin{figure*}[!th]
    \centering
    \includegraphics[width=\linewidth]{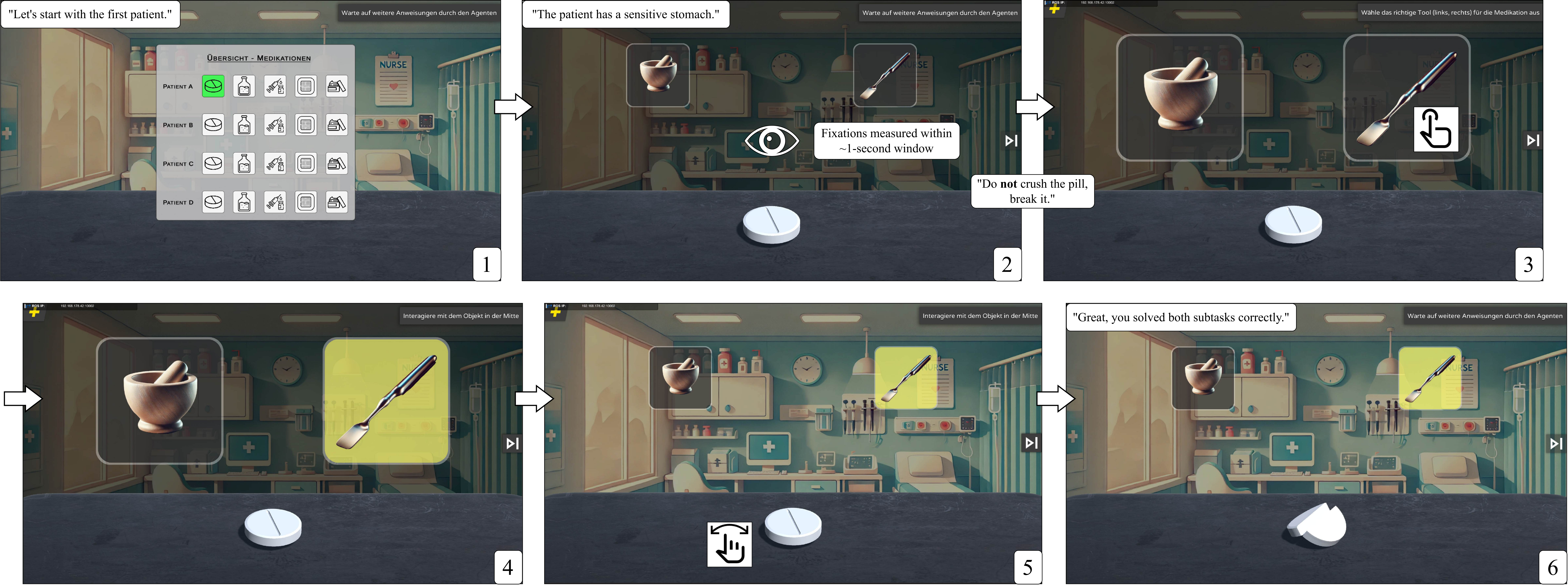}
    \caption{Overview of the task sequence, including: (1) the overarching goal of medication preparation, (2) verbal presentation of the patient's medical history and action instructions, (3-4) selection of the appropriate tool, (5) initiation of the interaction task and execution of the correct gesture, and (6) verbal feedback on task completion. Reaction times for both subtasks are measured at points (3) and (5), based on the first interaction with the touchscreen.}
    \label{fig:task_flow}
    \vspace{-18pt}
\end{figure*}

The experiment is divided into three phases.
First, participants' \acl{stm} is tested by presenting them with 10 words, which they are then asked to repeat.
Second, during the interactive part of the experiment, participants perform tasks on the touchscreen autonomously with the robot, without the experimenter present in the room.
The robot provides full guidance throughout the scenario.
This includes a short tutorial to learn gestures on the touchscreen and combine them with verbal actions.
Preparing medication for fictional patients is the main task of the participants.
Therefore, four imaginary patients, each with a different medical history, are introduced into the scenario (\autoref{tab:tasks}).
We developed two clinical stories focusing on the need for specific medication preparation for each task.
For example, \textit{Patient A} has a sensitive stomach and needs their medication to be broken up with a spatula.
Meanwhile, \textit{Patient B} is at risk of choking and requires the pill to be crushed with a mortar.
Participants are given 20 tasks (5 tasks, 4 patients).
The tasks are given sequentially, with e.g., all pills prepared for the four patients before moving on to the next medication.
By including two different solutions (tools) in the tasks, we can use explanatory strategies to manage attention between the two goals (\autoref{fig:tasks}).
By deciding the order of tool selection, we create scenarios that require shifting attention between goals.
Five different presentation patterns are defined for the tasks, which determine the order in which participants should select the correct tool and interact with the corresponding gesture.
These patterns include alternating (2, 1, 2, 1), paired (1, 1, 2, 2), hugging (1, 2, 2, 1), biased (1, 1, 1, 2), and converging~(2, 2, 1,~2) arrangements, each varying the distribution and repetition of tool selections across iterations.
Each participant follows a predetermined sequence of task presentations with their patterns.
The task sequence (\autoref{fig:task_flow}) follows a structured flow: 
(1)~the participant is given an overview of the experiment, including the overall goal of the study; 
(2)~they are given verbal information about the patient's condition, followed by a brief pause to allow them to reflect and draw conclusions from the diagnosis;
(3)~instructions are given with an explanatory strategy describing the correct tool and its intended action, after which the participant must select the appropriate tool; 
(4)~once selected, the tool is highlighted and the interaction phase is activated; 
(5)~the participant performs the required gesture to modify the target object;
(6)~at the end of the interaction, they receive verbal feedback on both subtasks.
After completing the interactive tasks with the NAO, participants will be asked to complete a short online questionnaire that collects demographic information, assesses technology affinity, and gathers feedback on subjective user experience and recall.

\subsection{Monitoring and Measurements} \label{sec:measurements}

\textbf{Monitoring}: For the adaptive selection of scaffolding strategies in the group \textit{SHIFT}, it is essential to extract social cues from the interaction between the human and the robot.
These cues are then used to generate adaptive strategies that are tailored to the individual’s specific needs.
For \textit{SHIFT}~\cite{gross2025shift}, we focus on collecting data about \textbf{(I)}~\ac{vfoa}, \textbf{(II)}~task performance, and \textbf{(III)}~scaffolding strategy history.
These inputs allow \textit{SHIFT} to assess the participant's cognitive state based on its definitions of gaze distribution, task awareness, and processing capacity \cite{gross2025shift}, enabling \textit{SHIFT} to recommend an appropriate scaffolding strategy (\autoref{tab:verbal_instructions}).
\textbf{(I)}~We record fixations on four \ac{aoi} in the study setting (tool 1, tool~2, object, and NAO).
It is crucial to note that we are not assessing a global level of attention, such as engagement, but rather the \ac{vfoa} within a one-second window after the verbal presentation of the patient's medical history and before the instruction of the action to be selected in the selection task (\autoref{fig:task_flow}: transition from scene 2 to scene~3).
This approach allows us to specifically capture the gaze behavior related to the participant's intentions and to estimate whether the participant knows which tool to choose based on the medical history alone.
If the gaze data reveals uncertainty or a preference for the wrong tool, \textit{SHIFT} can use a targeted explanation strategy to redirect attention to the correct tool.
\textbf{(II)}~We track the success and time spent on each subtask independently and input this data into \textit{SHIFT}, which calculates a task performance score.
To assess true task understanding beyond performance, we evaluate the subtasks separately: the selection task reflects understanding at the comprehension level, while the interaction task measures understanding at the enabledness level \cite{buschmeier2023formsunderstandingxaiexplanations}.
This approach enables \textit{SHIFT} to differentiate between misconceptions in understanding and determine the most appropriate scaffolding strategy to apply.
\textbf{(III)}~The task performance data also incorporates the explanatory strategy applied, from which \textit{SHIFT} estimates the participant’s processing capacity based on the cognitive demands of the strategy.

\textbf{Measurements}: The goal of the study is to evaluate participants' cognitive processing costs when engaging with different explanatory strategies, as well as the impact of these strategies on task understanding.
Different scaffolding strategies require varying levels of cognitive effort to process verbal input.
These differences are seen in reaction times, or the time taken to respond after receiving a verbal strategy.
We define \textbf{processing costs} as the reaction time in task solving.
The reaction time for the selection task, measured from when a verbal scaffolding strategy is delivered (verbal utterance ends) until the first interaction with the touchscreen (\autoref{fig:task_flow}, image 3).
For the interaction task, reaction time is recorded from the moment the selection task ends (tools are zoomed out) until the first interaction with the touchscreen (\autoref{fig:task_flow}, image 5).
To quantify this, we average the reaction times across both subtasks (selection and interaction).
\textbf{Task understanding} was assessed by tracking errors as an indicator of the effectiveness of an explanation strategy.
The total number of errors for each trial was summed up from both subtasks (selection and interaction) across all tasks, iterations, and participants.
An error is defined as: Solving a task incorrectly or completing a task prematurely.
Following the study, participants completed a \textit{questionnaire} via SoSci Survey \cite{soscisurvey}, which collected demographic data, the ATI \cite{ati}, and subjective task difficulties with the \ac{seq} \cite{sauro2009comparison}.
A \ac{stm} test was conducted, which involved a word repetition task before the interactive part of the study.

\subsection{Technical Setup}

Throughout the study, video recordings capture both the screen and the interaction between the human and the robot.
Eye movements were recorded using the Pupil Core eye tracker from Pupil Labs \cite{pupil_core_2014}, with a 5-point calibration procedure.
The study scenario, a touchscreen game, was developed using Unity3D and allows for full automation through a \ac{scxml}-based approach.
The 3D models in the scenario were created using Blender, while the background images and tool graphics were generated with ChatGPT and DALL-E 3 \cite{dalle}.
Communication between the robot, scenario, and \textit{SHIFT} is facilitated via RISE \cite{gross_schuetze_rise}, a system based on \ac{ros} \cite{ros} designed to support the implementation of studies in a robotics context.

\section{Results}

This study evaluates the effects of scaffolding strategies on human processing costs (\autoref{sec:results_processing_costs}) and changes in task understanding (\autoref{sec:results_task_understanding}).
We evaluate the functionalities of \textit{SHIFT} by analyzing the classifications of its monitoring components, the cognitive states of the participants, and the selection of explanatory strategies based on observed human behavior during the experiments (\autoref{sec:model_evaluation}).
The analysis was performed in R \cite{rstudio24}.

\subsection{Effects on Processing Costs} \label{sec:results_processing_costs}

\begin{figure}[!th]
    \centering
    \includegraphics[width=\linewidth]{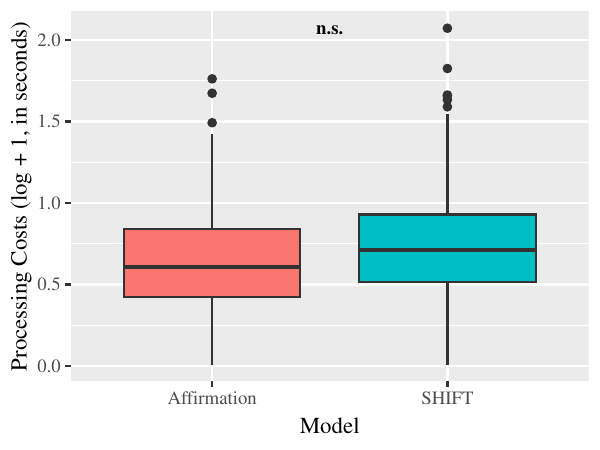}
    \vspace{-20pt}
    \caption{Time in seconds until the first interaction with the touchscreen averaged for selection and interaction task as processing costs. Processing costs for experiment running with \textit{SHIFT} and with affirmations (\textit{BL}).}
    \label{fig:processing_cost}
    \vspace{-10pt}
\end{figure}

\autoref{fig:processing_cost} describes the processing costs for the groups \textit{SHIFT} and baseline.
To reduce the effect of outliers and to improve the presentation of the data, we apply a logarithmic transformation ($\textit{log} + 1$) to normalize the processing costs.
A \ac{lme} was fitted using the lme4 package in R \cite{lmer} to analyse the effect of model use (\textit{SHIFT} vs. \textit{BL}) on processing cost while accounting for individual differences.
In this model, the participant IDs were included as a random intercept to capture baseline variability in processing cost, reflecting the fact that processing costs are nested within participants and observations from the same participant are not independent.
The results showed no significant effect by the use of \textit{SHIFT}, \mbox{$\beta = 0.118$, $SE = 0.059$, $t(28) = 1.99$, $p = 0.056$}, but suggesting a trend toward increased processing cost in the \textit{SHIFT} condition compared to the baseline.

\subsection{Effects on Task Understanding} \label{sec:results_task_understanding}

\begin{figure*}[!t]
    \centering
    \includegraphics[width=\linewidth]{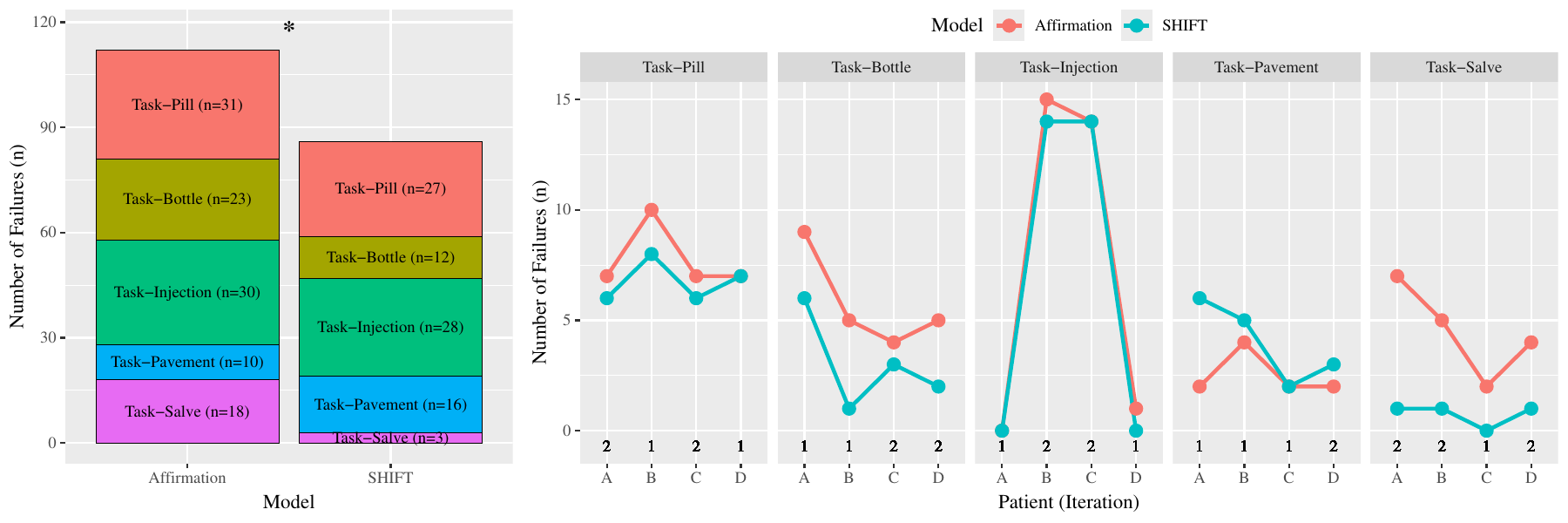}
    \vspace{-20pt}
    \caption{Evaluation of the task understanding by task failure rate. \textbf{Left}: Comparison of task failures as error-rates for \textit{SHIFT} and baseline. In baseline condition, the verbal instruction is always an affirmation. With the use of \textit{SHIFT}, the strategies are selected by the observation of the human cognitive state. \textbf{Right}: The patient (iterations) describes the number of repetition in a task, each task is repeated 4 times. Visualization of the changes in the total task performance failure sum over time. The numbers at the bottom indicate the correct tool to be selected for task completion, serving as the target of discourse. The order of these targets follows specific patterns, including alternating (2, 1, 2, 1), paired (1, 1, 2, 2), hugging (1, 2, 2, 1), biased (1, 1, 1, 2), and converging (2, 2, 1, 2) arrangements. Each pattern varies in how the tool selections are distributed and repeated across iterations.}
    \label{fig:task_understanding_failures}
    \vspace{-18pt}
\end{figure*}

Task understanding is measured by the error rate in task completion.
In \autoref{fig:task_understanding_failures}, we compare the total number of errors between \textit{SHIFT} and baseline conditions.
The total number of failures is $n_\textit{BL} = 112$ and $n_\textit{SHIFT} = 86$, a reduction of $23.21\%$.
While the results showed that the overall effect of model usage (\textit{SHIFT} vs. \textit{BL}) was not statistically significant ($p = 0.065$), our analysis focuses on the error rates for each group across different tasks, aiming to evaluate the interaction effects between task failures and tasks within each group.
We created a contingency table that examines the interaction of error rates with tasks for both groups.
A Fisher's Exact Test revealed a statistically significant difference ($p = 0.011$), indicating that the error rate distributions between \textit{SHIFT} and baseline conditions vary depending on the task, with an interaction effect between model usage and task failures.
The relationship between model usage and task was evaluated using Cramér's V~\cite{cramer1946contribution}, which revealed a small effect size ($V = 0.253$).
For further evaluation of the interaction effect between model usage and task, we examined the total error rates per task across iterations.
To assess the influence of model usage over time, we analyzed the changes in task performance failure rates across repetitions (coded as trials with different patients), with the results summed up across all participants.
\autoref{fig:task_understanding_failures} presents that in four out of five tasks, the use of \textit{SHIFT} resulted in lower error rates compared to the baseline.
According to the \ac{seq} \cite{sauro2009comparison} measured on a 7-point Likert scale, \textit{Task-Pill} ($\textit{SEQ} = 3.76$) was rated as the most difficult on average across all participants, followed by \textit{Task-Injection} ($\textit{SEQ} = 3.74$), \textit{Task-Salve} ($\textit{SEQ} = 2.52$), and \textit{Task-Bottle} ($\textit{SEQ} = 2.45$), with \textit{Task-Pavement} ($\textit{SEQ} = 2.09$) being rated as the easiest.

\subsection{Model Evaluation} \label{sec:model_evaluation}

\begin{figure*}[!t]
    \centering
    \includegraphics[width=\linewidth]{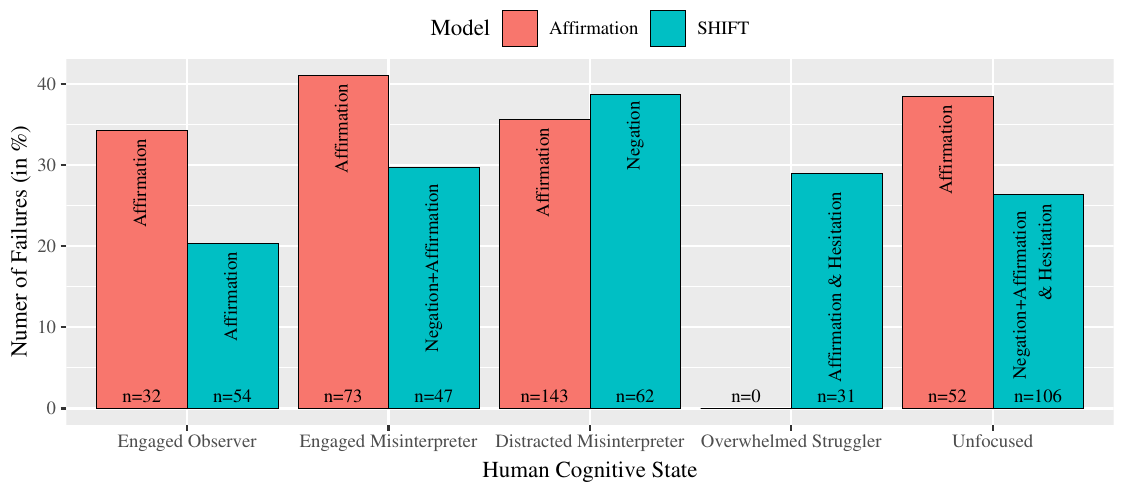}
    \vspace{-24pt}
    \caption{Evaluation of cognitive states based on processing capacity, gaze distribution, and task awareness as defined by \textit{SHIFT}. The percentage of failures in each cognitive state is reported for \textit{SHIFT} and the baseline, with the total number of state visits (n) indicated at the bottom of each bar.}
    \label{fig:cognitive_states_evaluation}
    \vspace{-18pt}
\end{figure*}

We evaluate task error rates in relation to the participant's current level of understanding.
This assessment is based on gaze distribution, processing capacity, and task awareness during the human-robot-task interaction.
These factors define the human cognitive state within \textit{SHIFT}.
Using this cognitive state, \textit{SHIFT} adaptively selects the most appropriate scaffolding strategy.
\autoref{fig:cognitive_states_evaluation} visualizes the error rate in percent relative to each participant's cognitive state throughout the experiment.
The results show that in three of the five cognitive states -- \textit{Engaged Observer} (34\% \textit{BL} vs. 20\% \textit{SHIFT}), \textit{Engaged Misinterpreter} (41\% vs. 29\%) and \textit{Unfocused} (38\% vs. 26\%) -- \textit{SHIFT} reduces the error rate compared to the baseline.
In the \textit{Distracted Misinterpreter} (35\% vs. 38\%) state, \textit{SHIFT} uses negation as the optimal explanation strategy.
This approach leads to a higher error rate than in the baseline, where a simple affirmation would be more beneficial.
In the current implementation of \textit{SHIFT}, processing capacity is not classified as ,,low" within 20 tasks when the explanatory strategy remains consistent (affirmation only in the baseline).
Consequently, the baseline does not reach the \textit{Overwhelmed Struggler} (0\% vs. 29\%) state.

\section{Discussion}

The evaluation of this study focuses on processing costs and task understanding in robot-guided task solving (\autoref{sec:research_questions}) for addressing \textbf{research question 1}.

$H_1$) Based on previous research \cite{gros_scaffolding_nodate, banh2024benefiting}, more cognitively demanding verbal scaffolding strategies, such as negation, tend to increase human processing costs compared to simpler affirmations.
Although not statistically significant, the results (\autoref{fig:processing_cost}) suggest a clear trend toward increased processing costs, as indicated by longer reaction times when using scaffolding strategies adaptively generated by \textit{SHIFT}~\cite{gross2025shift} compared to the baseline condition with only affirmations.
While the experimental design and current sample size preclude definitive statistical confirmation of our hypothesis, our results are consistent with its assumptions and provide supporting evidence for communication in \ac{hri}, even when \textit{SHIFT} is used to generate verbal scaffolding strategies.

$H_2$) In addition to increased processing costs, more complex scaffolding strategies can also benefit participants.
Previous studies \cite{gros_scaffolding_nodate} have suggested that negations improve action understanding, and our study confirms that in certain contexts -- particularly when a task is more difficult -- the adaptive scaffolding strategies provided by \textit{SHIFT} outperform pure affirmations (\autoref{fig:task_understanding_failures}).
The results show that the total number of failures is reduced with the use of \textit{SHIFT}.
The number of failures is lower compared to the baseline in 4 out of 5 tasks .
Only for the \textit{Task-Pavement}, \textit{SHIFT} performs worse than the baseline.
According to the results from the subjective ratings of the questionnaire, the \textit{Task-Pavement} is rated as the easiest task.
This could mean that more extensive scaffolding strategies only make sense for tasks that are not easy to solve and where the explanation actually provides additional information.
Scaffolding for tasks that are too easy could therefore be overwhelming.
This aligns with existing literature \cite{banh2024benefiting}, which suggests that negations, for example, are only beneficial when there is something to negate, such as changing expectations or breaking a pragmatic frame \cite{rohlfing2016alternative}.
A similar effect applies to hesitations. 
\cite{richter2021attention} demonstrated that in particular, individuals with poor memory performance (for whom recall tasks are more challenging) benefit from hesitations.
Our results provide support for our hypotheses: 
Adaptive scaffolding based on the human cognitive state tends to increased processing costs (\autoref{fig:processing_cost}), but also foster task understanding, as demonstrated by a lower error rate when using \textit{SHIFT} (\autoref{fig:task_understanding_failures}).

For answering \textbf{research question 2}, we evaluated error rates in relation to participants' cognitive states while considering the role of negation in explanatory strategies.
As illustrated in \autoref{fig:cognitive_states_evaluation}, \textit{SHIFT} demonstrates a lower percentage error rate in three out of five cognitive states compared to the baseline. 
However, in the \textit{Distracted Misinterpreter} state -- where \textit{SHIFT} only uses negation -- a pure affirmation strategy appears to produce fewer errors.
Negation is known to inhibit cognitive processing.
Negative sentences take longer to process and lead to higher error rates than their affirmative counterparts, especially when presented out of context \cite{beltran2021inhibitory}.
This suggests that while negation can be beneficial when there is a clear contextual need for negation, it can also impose additional processing costs that may hinder performance if not applied appropriately.
Furthermore, our task design may not have adequately simulated the counterfactual scenarios necessary for optimal negation processing.
Overall, these findings suggest that while the \textit{SHIFT} scoring system shows promise, further refinement of its negation-based strategies is needed to fully exploit its potential for improving task understanding.
The challenge in designing \ac{hri} studies lies in balancing task complexity and creating scenarios where explanatory strategies are both effective and reflect natural, real-world situations.
Without sufficient complexity, the adaptability of \textit{SHIFT} is limited and its advantage over simple affirmations remains small.
Our findings suggest that not all tasks were sufficiently complex to fully engage the adaptive scaffolding strategies. 
Furthermore, because the tasks were completed independently and did not build on each other, the potential of \textit{SHIFT} -- which relies on interaction history to determine the appropriate scaffolding strategy -- was not fully exploited.
To better assess the \textit{SHIFT}'s adaptability and the differences between scaffolding strategies directly, future studies should include tasks of greater complexity and a sequential design.
A further improvement of the adaptation strategy might be achieved by a learning strategy that could, for example, learn that a pure negation strategy as administered by \textit{SHIFT} does lead to errors and change toward a more successful strategy.
However, in a prior study of \textit{SHIFT} with synthetic data and a \acl{rl} approach we could show that in order to achieve scaffolding strategy that is better than our hand-crafted model more than 50 iterations are needed \cite{gross2025shift}.
Therefore, further strategies are required to increase \textit{SHIFT}'s performance before learning becomes a viable alternative.
Moreover, \textit{SHIFT} does not adaptively change the content of the strategy; it dynamically selects the type of explanation.
The interaction between ,,what" is explained and ,,how" it is explained \cite{klein2009erklaren} is also an exciting avenue for future research.
In addition, the interaction effects between hesitations and negations need further investigation.

\section{Conclusion} 

This research improves our understanding of how different scaffolding strategies affect processing costs and task understanding in \acl{hri}.
The results highlight the importance of adaptively tailoring explanations based on a participant's cognitive state in specific, different task situations.
Using different scaffolding strategies tends to increases of the cost of processing, resulting in higher cognitive load and additional processing loops.
However, the additional processing costs associated with more complex scaffolding strategies can have a positive impact on task understanding in situations that require a shift in attention and modification of established expectations.
Furthermore, our results show that not all scaffolding strategies are generally effective.
For example, negation should be used selectively, as its effectiveness depends on the participant's cognitive state and the task context.
These findings emphasize that scaffolding strategies should be used specifically in connection with the complexity of tasks.
We showed how dynamic adaptation can reduce errors and improve task understanding by examining \textit{SHIFT}, which adapts scaffolding strategies based on human cognitive states.
The study highlights the relationship between cognitive load (processing cost) and task understanding (task performance), and provides a foundation for fields such as \ac{xai}, robotics and cognitive science to develop more personalized and context-aware robotic systems.


%%%%%%%%%%%%%%%%%%%%%%%%%%%%%%%%%%%%%%%%%%%%%%%%%%%%%%%%%%%%%%%%%%%%%%%%%%%%%%%%

%%%%%%%%%%%%%%%%%%%%%%%%%%%%%%%%%%%%%%%%%%%%%%%%%%%%%%%%%%%%%%%%%%%%%%%%%%%%%%%%
%\section*{APPENDIX}

%\vspace*{-2\baselineskip}
%\IEEEtriggercmd{\enlargethispage{-5.35in}}
%\IEEEtriggeratref{40}
\bibliographystyle{IEEEtran}
\bibliography{references}

\end{document}